\documentclass[preprint]{revtex4}

\usepackage{graphicx}
\usepackage{dcolumn}
\usepackage{amsmath}

\usepackage{array}
\usepackage{rotating}

\usepackage{epstopdf} 

\usepackage{latexsym}

\usepackage{float}

\restylefloat{figure}

\renewcommand\thesection{\arabic{section}}

\makeatletter
\def\p@subsection{}
\makeatother
\def\p@subsection{\thesection\,}
\allowdisplaybreaks

\begin{document}
\title{New charged anisotropic compact models}
\author{D. Kileba Matondo}\email[]{dkilebamat@gmail.com}
\author{S. D. Maharaj}\email[]{maharaj@ukzn.ac.za}
\affiliation{Astrophysics and Cosmology Research Unit, School of Mathematics, Statistics and Computer Science, 
University of KwaZulu--Natal, Private Bag X54001, Durban 4000, South Africa}

\begin{abstract}
\noindent We find new exact solutions to the Einstein-Maxwell field equations which are relevant in the description of highly compact stellar objects. The relativistic star is charged and anisotropic with a quark equation of state. Exact solutions of the field equations are found in terms of elementary functions. It is interesting to note that we regain earlier quark models with uncharged and charged matter distributions. A physical analysis indicates that the matter distributions are well behaved and regular throughout the stellar structure. A range of stellar masses are generated for particular parameter values in the electric field. In particular the observed mass for a binary pulsar is regained.\\

\noindent
\end{abstract}


\maketitle

\section{Introduction}
To obtain an understanding of the gravitational dynamics of a general relativistic star it is necessary to solve the Einstein-Maxwell equations. The matter distribution may be anisotropic in the presence of an electromagnetic field. On physical grounds we should include an equation of state relating the radial pressure to the energy density in a barotropic distribution. In this way we can model relativistic compact objects including dark energy stars, quark stars, gravastars, neutron stars and ultradense matter. For some recent models investigating the properties of charged anisotropic stars see the treatments of Feroze and Siddiqui \cite{hh1}, Thirukkanesh and Ragel \cite{hh2}, Maurya and Gupta \cite{hh3}, Maurya \textit{et al.} \cite{hh4}, Pandya \textit{et al.} \cite{hh5}, Bhar \textit{et al.} \cite{hh6} and Murad \cite{hh7}.

These exact solutions of the Einstein-Maxwell system are essential for describing astrophysical processes. Our study in this paper is related to describing the stellar interior with a spherically symmetric anisotropic charged matter distribution. We impose a quark equation of state on the model. The physical features of quark stars have been recently investigated by Malaver \cite{hh8}, Mafa Takisa \textit{et al.} \cite{hh9}, Sunzu \textit{et al.} \cite{hh10} and Paul \textit{et al.} \cite{hh11}. The investigations of Sharma \textit{et al.} \cite{hh12} show that the redshift, luminosity and maximum mass of a compact body are affected by the electric field and anisotropy. The works of Mak and Harko \cite{hh13}, Komathiraj and Maharaj  \cite{hh14,hh15} and Maharaj and Komathiraj \cite{hh16} show the significant role of the electromagnetic field in describing the gravitational behaviour of  compact stars composed of quark matter. The role of anisotropy has been highlighted by Dev and Gleiser  \cite{hh17,hh18}. By making a specific choice of the electric field, Hansraj and Maharaj  \cite{hh19} obtained solutions with isotropic pressures to the Einstein-Maxwell system. These solutions satisfy a barotropic equation of state and contain the Finch and Skea \cite{hh20} model. Charged anisotropic models with a linear equation of state were found by Thirukkanesh and Maharaj \cite{hh21}. Mafa Takisa and Maharaj \cite{hh22} utilised a linear equation of state to generate regular solutions of anisotropic spherically symmetric charged distributions which can be related to observed astronomical objects. 

The main objective of this paper is to generate a new class of exact solutions to the Einstein-Maxwell system of equations that is physically acceptable. We generalize the solution to the Einstein-Maxwell system, with no singularity in the charge distribution at the centre, obtained by Mafa Takisa and Maharaj \cite{hh22} by generating a new class of exact solutions. The physical features such as the gravitational potentials, electric field intensity, charge distribution and matter distribution are well behaved. These new exact solutions describe a charged relativistic sphere with anisotropic pressures and a linear quark equation of state. In Sect.~\ref{sec2}, we give the expression of the line element modelling the interior of relativistic star which allows us to rewrite the Einstein-Maxwell equations in terms of new variables. In Sect.~\ref{sec3}, we present the basic assumptions for the physical models. We make a choice for one of the gravitational potentials and the electric field intensity which allow us to integrate the field equations. Three new classes of exact solutions to the Einstein-Maxwell field equations, in terms of elementary functions, are determined in Sect.~\ref{sec3}. The first class is unphysical and the other two classes satisfy the physical criteria. In Sect.~\ref{sec4}, we demonstrate how our exact solutions regain the uncharged anisotropic solution found by Thirukkanesh and Maharaj \cite{hh21}, and the charged anisotropic solution found by Mafa Takisa and Maharaj \cite{hh22}. In Sect.~\ref{sec5}, we study the physical features of our models and present graphs for the energy density, radial pressure, electric field intensity, charge density and mass. The graphs indicate that the matter and electromagnetic quantities are well behaved. In Sect.~\ref{sec6}, we generate stellar masses and show that our charged general relativistic solutions can be related to observed astronomical objects. Concluding comments are made in Sect.~\ref{sec7}.

\section{Field equations\label{sec2}}
The line element for static spherically symmetric spacetimes, modelling the interior of the relativistic star, has the form
\begin{equation}
\label{A1}
ds^2 = -e^{2\nu(r)}dt^{2}+e^{2\lambda(r)}dr^{2}+r^{2}(d\theta^{2}+\sin^{2}\theta d\phi^{2}).
\end{equation}
The functions $\lambda(r)$ and $\nu(r)$ correspond to the gravitational potentials. The energy momentum tensor ${\bf T}$ therefore has the general form
\begin{eqnarray}
\label{A2}
T_{ab}&=&\mbox{diag}(-\rho-\frac{1}{2}E^{2}, p_{r} -\frac{1}{2}E^{2},\nonumber\\
  && p_{t} +\frac{1}{2}E^{2}, p_{t} +\frac{1}{2}E^{2}),
\end{eqnarray}
where the quantities $\rho, p_{r}, p_{t}$ and $E$ are respectively, energy density, radial pressure, tangential pressure and electric field intensity. Then the Einstein-Maxwell system of equations can be written in the form
\begin{subequations} 
\label{A3}
\begin{eqnarray}
\label{A3a}
\frac{1}{r^2}\left[r(1-e^{-2\lambda})\right]'&=&\rho+\frac{1}{2}E^2,\\
\label{A3b}
-\frac{1}{r^2}(1-e^{-2\lambda}) + \frac{2\nu'}{r}e^{-2\lambda} &=& p_{r}-\frac{1}{2}E^2,\\
\label{A3c}
e^{-2\lambda}\left(\nu'' +\nu'^{2} + \frac{\nu'}{r} - \nu'\lambda' -\frac{\lambda'}{r}\right) &=& p_{t}+\frac{1}{2}E^2,\\
\label{A3d} 
\sigma=\frac{1}{r^2}e^{-\lambda}(r^{2}E)',
\end{eqnarray}
\end{subequations}
in terms of the coordinate $r$. An equivalent form of the Einstein-Maxwell field equations is obtained if we introduce the transformation 
\begin{eqnarray}
\label{A4}
x = Cr^{2},~~ Z(x) = e^{-2\lambda (r)},~~ A^{2}y^{2}(x) = e^{2\nu(r)},
\end{eqnarray}
where $A$ and $C$ are constants. This transformation was first used by Durgapal and Bannerji \cite{hh23}. The line element ($\ref{A1}$) then has the form
\begin{eqnarray}
\label{A5}
ds^{2}&=&-A^{2}y^{2}(x)dt^{2}+\frac{1}{4CxZ(x)}dx^{2}\nonumber\\
  && +\frac{x}{C}\left(d\theta^{2}+\sin^{2}\theta d\phi^{2}\right),
\end{eqnarray}
in terms of the variable $x$. The field equations $(\ref{A3})$ become
\begin{subequations} 
\label{A6}
\begin{eqnarray}
\label{A6a} 
\frac{\rho}{C}&=& -2\dot{Z}+\frac{1-Z}{x}-\frac{\ E^2}{2C},\\
\label{A6b} 
\frac{ p_{r}}{C}&=&4Z\frac{\dot{y} }{y}+\frac{Z-1}{x}+\frac{\ E^2}{2C},\\
\label{A6c} 
\frac{ p_{t}}{C}&=&4xZ\frac{\ddot{y}}{y}+(4Z+2x\dot{Z})\frac{\dot{y} }{y}+\dot{Z}-\frac{\ E^2}{2C},\\
\label{A6d}
\frac{\sigma^2}{C}&=&\frac{4Z}{x}(x\dot{E}+E)^2.
\end{eqnarray}
\end{subequations}
The mass of a gravitating object within the stellar radius, is important for comparison with observations. The mass contained within a radius $x$ of the sphere is given by the expression
\begin{equation}
\label{A7}
M(x)=\frac{1}{4C^{3/2}}\int^{x}_{0}\sqrt{\omega}\rho(\omega)\ d\omega.
\end{equation}
This expression is sometimes called the mass function. For a physically realistic relativistic star, we expect that the matter distribution should obey a barotropic equation of state of the form $p_r=p_r(\rho)$. For a charged anisotropic matter distribution we consider the linear relationship
\begin{equation}
\label{A8}
p_{r} = \alpha\rho-\beta,
\end{equation}
where $\alpha$ and $\beta$ are constants.

Then the Einstein-Maxwell system governing the gravitational interactions of a charged anisotropic body, with a linear equation of state, can be written as
\begin{subequations} 
\label{A9}
\begin{eqnarray}
\label{A9a}
\frac{\rho}{C}&=&\frac{1-Z}{x}-2\dot{Z}-\frac{\ E^2}{2C},\\
\label{A9b}
p_{r}&=&\alpha \rho-\beta,\\
\label{A9c}
p_{t}&=&p_{r}+\Delta,\\
\label{A9d}
\Delta&=&4CxZ\frac{\ddot{y}}{y} + 2C\left(x\dot{Z}+\frac{4Z}{1+\alpha}\right)\frac{\dot{y}}{y} \nonumber\\ & & +\left(\frac{1+5\alpha}{1+\alpha}\right)C\dot{Z} - \frac{C(1-Z)}{x} + \frac{2\beta}{1+\alpha},\\
\label{A9e}
\frac{E^2}{2C}&=&\frac{1-Z}{x} - \left(\frac{1}{1+\alpha}\right)\left(2\alpha\dot{Z}+4Z\frac{\dot{y}}{y} +\frac{\beta}{C}\right),\\
\label{A9f}
\frac{\sigma^2}{C}&=&\frac{4Z}{x}(x\dot{E}+E)^2.
\end{eqnarray}
\end{subequations}
The quantity $\Delta = p_{t} - p_{r}$ is defined as the measure of anisotropy and vanishes for isotropic pressures.
The system of equations $(\ref{A9})$ is nonlinear consisting of six equations with eight independent variables $y$, $Z$, $\rho$, $p_{r}$, $p_{t}$, $E$, $\sigma$ and $\Delta$. We need to select two of the variables involved in the integration process to solve the system $(\ref{A9})$.

\section{Physical models\label{sec3}}
To generate a new class of solutions to the Einstein-Maxwell system requires a choice for one of the gravitational potentials and the electric field intensity. In our approach we make the particular choice
\begin{subequations}
\label{A10}
\begin{eqnarray}
\label{A10a}
Z&=&\frac{1+\left(a-b\right)x}{1+ax},\\
\label{A10b}
\frac{E^2}{C}&=&\frac{l a^{3}x^{3} + s a^{2}x^{2} + k(3+ax)}{(1+ax)^2},
\end{eqnarray}
\end{subequations}
where $a$, $b$, $s$, $k$, $l$ are real constants. It is important to note that we keep the same form for the gravitational potential first used by Thirukkanesh and Maharaj \cite{hh21}. The function $Z$ is finite at the centre $x=0$ and regular in the interior. The function for the electrical field $E$ is a generalised form that contains previous studies as special cases. The function $E$ is finite at the centre and well behaved in the stellar interior. When $l=s=0$ we regain the models of Thirukkanesh and Maharaj \cite{hh21}. If $l=0$ then the exact models of Mafa Takisa and Maharaj \cite{hh22} are obtained. For particular choices of the parameters $a$ and $b$ we regain other earlier models: charged Hansraj and Maharaj \cite{hh19} stars $(a=b=1)$, charged Maharaj and Komathiraj \cite{hh16} stars $(a=1)$, uncharged Finch and Skea \cite{hh20} stars $(a=b=1)$, uncharged Durgapal and Bannerji \cite{hh23} neutron stars $(a=1, b=-3/2)$, and uncharged Tikekar \cite{hh24} superdense stars $(a=7, b=8)$.

From ($\ref{A10}$) and ($\ref{A9e}$) we obtain the first order equation
\begin{eqnarray}
\label{A11}
\frac{\dot{y}}{y}&=&\frac{4\alpha b -(1+\alpha)(la^{3}x^{3} +sa^{2}x^{2} +k(3+ax))}{8(1+ax)\left[1+(a-b)x\right]}\nonumber\\& & -\frac{\beta(1+ax)}{4C\left[1+(a-b)x\right]} +\frac{(1+\alpha)b}{4\left[1+(a-b)x\right]}.
\end{eqnarray}
It is necessary to integrate ($\ref{A11}$) to complete the model of a charged gravitating sphere. It is convenient to categorise our solutions in terms of the constant $b$. We consider, in turn, the following three cases: $b=0$, $b=a$, $b\neq a$.

\subsection{The case $b=0$}
When $b=0$, equation ($\ref{A11}$) assumes the simple form 
\begin{eqnarray}
\label{A12}
\frac{\dot{y}}{y}&=&\frac{(1+\alpha)(la^{3}x^{3} +sa^{2}x^{2} + k(3+ax))}{-8\left(1+ax\right)^2}\nonumber\\& &
  -\frac{\beta}{4C}.
\end{eqnarray}
This gives after integration the solution
\begin{eqnarray}
\label{A13}
y&=&D\left(1+ax\right)^{(-(3l+k-2s))(1+\alpha)/(8a)}\nonumber\\ & & \times
\exp\left[F(x)\right],
\end{eqnarray}
where $D$ is a constant of integration and 
\begin{eqnarray}
\label{A14}
F(x)&=&\frac{(1+\alpha)}{16(1+ax)}\nonumber\\ & & \times \left[4k-2asx(2+ax) \right.\nonumber\\ & & \left.
 + l(3+9ax+3a^{2}x^{2}-a^{3}x^{3})\right] \nonumber\\& & -\frac{\beta x}{4C}.
\end{eqnarray}
The energy density $\rho=-\frac{\ E^2}{2}$ generated by the potential $y$ in ($\ref{A13}$) is negative and the model is unphysical.

\subsection{The case $b=a$ \label{f1} }
\label{The case $b=a$}
When $b = a$, the differential equation ($\ref{A11}$) yields the form
\begin{eqnarray}
\label{A15}
\frac{\dot{y}}{y}&=&\frac{4\alpha a -(1+\alpha)(la^{3}x^{3} +sa^{2}x^{2} +k(3+ax))}{8(1+ax)}\nonumber\\& 
& -\frac{\beta(1+ax)}{4C} +\frac{(1+\alpha)a}{4}.
\end{eqnarray}
This equation has solution 
\begin{eqnarray}
\label{A16}
y&=&D\left(1+ax\right)^{(4\alpha a - (s+2k)(1+\alpha) +l(1+\alpha))/(8a)} \nonumber\\ & & \times
\exp\left[G(x)\right],
\end{eqnarray}
where $D$ is a constant of integration and the function $G(x)$ is given by
\begin{eqnarray}
\label{A17}
G(x)&=&-\frac{x}{16C} \left[2C(s-k)(1+\alpha) \right.\nonumber\\ & & \left.
 - a(c(sx-4)(1+\alpha)+2\beta x) -4\beta\right]x^{2} \nonumber\\& & -\frac{l(1+\alpha)}{48ac}\left(11+6ax-3a^{2}x^{2}+2a^{3}x^{3}\right).
\end{eqnarray}
The energy density $\rho$ generated by the potential $y$ in $(\ref{A16})$ is nonnegative. Then the complete solution to the Einstein-Maxwell field equations $(\ref{A9})$ can be written as
\begin{subequations}
\label{A18}
\begin{eqnarray}
\label{A18a}
e^{2\lambda}&=&1+ax,\\
\label{A18b}
e^{2\nu}&=&A^{2}D^{2}\left(1+ax\right)^{(4\alpha a - (s+2k)(1+\alpha) +l(1+\alpha))/(4a)} \nonumber\\ & & \times \exp\left[2G(x)\right],\\
\label{A18c}
\frac{\rho}{C}&=&\frac{-la^{3}x^{3}-sa^{2}x^{2}+(2a-k)(3+ax)}{2(1+ax)^2},\\
\label{A18d}
p_{r}&=&\alpha\rho-\beta,\\
\label{A18e}
p_{t}&=&p_{r}+\Delta,\\
\label{A18f}
\Delta&=&\frac{1}{16 c (1 + a x)^3}\nonumber\\ & & \times
 \left[C^{2}(k^{2}x(3 + a x)^{2}(1 + \alpha)^{2} \right.\nonumber\\ & & \left.
 + a^{2}x(a^{4}l^{2}x^{6}(1 + \alpha)^{2}\right.\nonumber\\ & & \left.
  + 2a^{3}lx^{4}(-2 + s x)(1 + \alpha)^{2} \right.\nonumber\\ & & \left.
  - 4(-3 + (8 - 9\alpha)\alpha + 4sx(2 + \alpha)) \right.\nonumber\\ & & \left.
  - 4 ax(2lx(5 + 3\alpha) - 2(2 + 3 \alpha(1 + \alpha))\right.\nonumber\\ & & \left.
	+ sx(6 + \alpha(5 + 3\alpha))) \right.\nonumber\\ & & \left.
	+ a^{2} x^{2} ((-2 + sx)^{2} (1 + \alpha)^{2} \right.\nonumber\\ & & \left.
	- 4lx(8 + \alpha(7 + 3 \alpha)))) \right.\nonumber\\ & & \left.
	+ 2k(-24 + ax(a^{3}lx^{4}(1 + \alpha)^{2} \right.\nonumber\\ & & \left.
	+ a^{2}x^{2}(-2 + 3lx + sx)(1 + \alpha)^{2} \right.\nonumber\\ & & \left.
	- 2(12 + \alpha(5 + 9 \alpha)) \right.\nonumber\\ & & \left.
	+ ax(3sx(1 + \alpha)^{2} \right.\nonumber\\ & & \left.
	- 2(7 + 9\alpha + 6 \alpha^{2}))))) \right.\nonumber\\ & & \left.
	+ 4cx(1 + ax)^{2}(3k(1 + \alpha) \right.\nonumber\\ & & \left.
	+ a^{3}lx^{3}(1 + \alpha) \right.\nonumber\\ & & \left.
	+ a^{2}x(-2 + sx)(1 + \alpha) \right.\nonumber\\ & & \left.
	+ a(-4 - 6\alpha + kx(1 + \alpha)))\beta  \right.\nonumber\\ & & \left.
	+ 4x(1 + a x)^{4} \beta^{2}\right],\\
\label{A18g}
\frac{E^2}{C}&=&\frac{la^{3}x^{3}+sa^{2}x^{2}+k(3+ax)}{(1+ax)^2},\\
\label{A18h}
\frac{\sigma^{2}}{C}&=&\frac{c(1+ax)^{-5}}{x(k(3 + ax) + a^{2}x^{2}(s + alx))} \nonumber\\ & & \times
\left[k(6 + ax(3 + ax)) \right.\nonumber\\ & & \left. +  a^{2}x^{2}(2s(2 + ax) \right.\nonumber\\ & & \left. 
+ alx(5 + 3ax))\right]^{2}.
\end{eqnarray}
\end{subequations}
The solution $(\ref{A18})$ is an exact solution of the Einstein-Maxwell system when $a=b$. 

We can compute the mass function explicitly from $(\ref{A7})$. In this case we can write the mass function in terms of $x$ by 
\begin{eqnarray}
\label{A19}
M(x)&=&\frac{1}{8C^{3/2}} \nonumber\\ & & \times
 \left[\frac{l\left(-6a^{3}x^{3}+14a^{2}x^{2}-70ax-105\right)x^{1/2}}{15a(1+ax)}\right.\nonumber\\& & \left.  
- \frac{5(6a(k-2a)x)x^{1/2}}{15a(1+ax)} \right.\nonumber\\& & \left.
-\frac{s(-15-10ax+2a^{2}x^{2})x^{1/2}}{15a(1+ax)}
\right. \nonumber\\& & \left. +\left(\frac{7l-5s}{a^{3/2}}\right)\arctan(\sqrt{ax})\right].
\end{eqnarray}
We can regain the mass functions of Thirukkanesh and Maharaj \cite{hh21} $(l=s=0)$ and Mafa Takisa and Maharaj \cite{hh22} $(l=0)$ from $(\ref{A20})$ as particular cases in the presence of charge. Note that their models necessarily contain an inverse tangent function. In our model when
\begin{equation}
\label{A20}
l = \frac{5}{7}s,
\end{equation}
the inverse tangent function is not present in the expression for $m(x)$.

\subsection{The case $b \neq a$ \label{g1}}
\label{g1}
The most interesting case is when $b \neq a$. When $b \neq a$, the solution of ($\ref{A11}$) can be written in the form
\begin{equation}
\label{A21}
y = D(1+ax)^{m}
\left[1+(a-b)x\right]^{n}\exp\left[F(x)\right],
\end{equation}
where $D$ is a constant of integration. The constants $m$, $n$ and the function $F(x)$ are given by
\begin{eqnarray}
\label{A22}
n&=&-\frac{a^{3}l(1+\alpha)}{8b}  \nonumber\\& &
+\frac{1}{8Cb(a-b)^{2}}\left[a^{2}c((s+2K)(1+\alpha)-4b\alpha) 
\right. \nonumber\\& & \left. +abC( 2b(1+5\alpha)-5k(1+\alpha)) 
\right. \nonumber\\& & \left. + b^{2}( 3Ck(1+\alpha) -2bC(1+3x) +2\beta) \right], \nonumber\\
\label{A23}
m&=&\frac{4b\alpha-(s+2k)(1+\alpha) +l(1+\alpha)}{8b},\nonumber\\
\label{A24}
F(x)&=&-\frac{ax}{16C(a-b)^{2}}\nonumber\\ & & \times
\left[(2C(1+\alpha)( s(a-b) +l(b-2a) )) \right.\nonumber\\& & \left. + (4(a-b)\beta + aCl(1+\alpha)(a-b)x)\right].\nonumber
\end{eqnarray}
As for the case $a=b$, the energy density $\rho$ generated from ($\ref{A21}$) is nonnegative. Then, we can write the exact solution for the system ($\ref{A9}$) as
\begin{subequations}
\label{A25}
\begin{eqnarray}   
\label{A25a}
e^{2\lambda}&=&\frac{1+ax}{1+(a-b)x},\\
\label{A25b}
e^{2\nu}&=&A^{2}D^{2}[1+(a-b)x]^{2n}(1+ax)^{2m}\nonumber\\ & & \times
\exp\left[-\frac{ax(2C(1+\alpha)( s(a-b) +l(b-2a) ))}{8C(a-b)^{2}} \right.\nonumber\\& & \left.-\frac{ax (4(a-b)\beta + aCl(1+\alpha)(a-b)x)}{8C(a-b)^{2}}\right],\\
\label{A25c}
\frac{\rho}{C}&=&-\frac{(k-2b)(3+ax)+a^{2}x^{2}(s+alx)}{2(1+ax)^2},\\
\label{A25d}
p_{r}&=&\alpha\rho-\beta,\\
\label{A25e}
p_{t}&=&p_{r}+\Delta,\\
\label{A25f}
\Delta&=&\frac{Ck}{8(1+ax)^{3}(1+(a-b)x)}\nonumber\\   & & \times
\left[-24+x(a^{4}lx^{4}(1+\alpha)^{2} \right.\nonumber\\ & & \left.
  -6b(2+\alpha)(-1+3\alpha)x\right.\nonumber\\ & & \left.
 +a^{3}x^{3}(4(-1+\alpha)+(3l+s)(1+\alpha)^{2}x)\right.\nonumber\\ & & \left.
+a^{2}x^{2}(3sx(1+\alpha)^{2}+8(-2+3\alpha) \right.\nonumber\\ & & \left. 
-2bx(-1+\alpha(4+\alpha)))\right.\nonumber\\ & & \left.
+2ax(2(-9+5\alpha)+bx(1-3\alpha(7+2\alpha)))\right]\nonumber\\ & &
 + \frac{Cx}{16(1+ax)^{3}(1+(a-b)x)}\nonumber\\   & & \times 
\left[4b^{2}(3+ax(4+ax)+12\alpha \right.\nonumber\\ & & \left.
+ 6\alpha ax(5+ax)+(3+ax)^{2}\alpha^{2})\right.\nonumber\\ & & \left.
+a^{4}s^{2}x^{4}(1+\alpha)^{2}-8a^{3}lx^{2}(5+3\alpha)\right.\nonumber\\ & & \left.
+a^{4}lx^{3}(a^{2}lx^{3}(1+\alpha)^{2}-32(2+\alpha) \right.\nonumber\\ & & \left.
-8ax(3+\alpha))\right.\nonumber\\ & & \left.
+2a^{2}sx(-8(2+\alpha)+ax(-8(3+\alpha)\right.\nonumber\\ & & \left. 
+ax(-8+alx^{2}(1+\alpha)^{2})))\right.\nonumber\\ & & \left.
 -4ab(20\alpha+24\alpha ax)\right.\nonumber\\ & & \left.
-4a^{2}bsx^{2}(-6+\alpha+3\alpha^{2} \right.\nonumber\\ & & \left.
+ax(-1+\alpha)(3+\alpha))\right.\nonumber\\ & & \left.
-4a^{3}bx^{2}(4\alpha+lx(-8+\alpha(-1+3\alpha)\right.\nonumber\\ & & \left.
+ax(-5+\alpha^{2})))\right]\nonumber\\ & & 
 + \frac{\beta x}{4(1+ax)(1+(a-b)x)}\nonumber\\   & & \times
\left[(1+\alpha)(k(3+ax)+a^{2}x^{2}(s+alx))\right.\nonumber\\ & & \left.
 -2b(2+3\alpha+ax(1+\alpha))\right]\nonumber\\ & & +
\frac{x(1+ax)\beta^{2}}{C(1+(a-b)x)} \nonumber\\ & &  + 
\frac{k^{2}x(3+ax)^{2}(1+\alpha)^{2}}{16(1+ax)^{3}(1+(a-b)x)},\\
\label{A25g}
\frac{E^2}{C}&=&\frac{la^{3}x^{3}+sa^{2}x^{2}+k(3+ax)}{(1+ax)^2},\\
\label{A25h}
\frac{\sigma^{2}}{C}&=&\frac{(1 + (a - b)x)}{x(1 + ax)^{5}(k(3 + ax) + a^{2}x^{2}(s + alx))} \nonumber\\ & & \times
\left[k(6 + ax(3 + a x))  \right.\nonumber\\& & \left. + a^{2} x^{2}(2s(2 + ax) + alx(5 + 3ax))\right]^{2}.
\end{eqnarray}
\end{subequations}
This exact solution of the Einstein-Maxwell equations is similar in structure to that when $a=b$ in Sect.~\ref{f1}. However note that $(\ref{A25})$ is a new exact solution written completely in terms of elementary functions.

For this solution the mass function is given by 
\begin{eqnarray}
\label{A26}
M(x)&=&\frac{1}{8C^{3/2}}\nonumber\\ & & \times
\left[\frac{l\left(-6a^{3}x^{3}+14a^{2}x^{2}-70ax-105\right)x^{1/2}}{15a(1+ax)}\right.\nonumber\\& & \left.  
- \frac{5(6a(k-2b)x)x^{1/2}}{15a(1+ax)} \right.\nonumber\\& & \left.
- \frac{s(-15-10ax+2a^{2}x^{2})x^{1/2}}{15a(1+ax)}
\right. \nonumber\\& & \left. +\left(\frac{7l-5s}{a^{3/2}}\right)\arctan(\sqrt{ax})\right].
\end{eqnarray}
This expression contains previously studied mass functions in the presence of charge. As in the previous case note that when
\begin{equation}
\label{A27}
l = \frac{5}{7}s,
\end{equation}
the inverse tangent function is absent in the expression for $M(x)$.

\section{Known solutions\label{sec4}}
When we set $s=l=0$ in $(\ref{A10b})$, then the gravitational potentials become
\begin{subequations}
\label{A28}
\begin{eqnarray}   
\label{A28a}
e^{2\lambda}&=&\frac{1+ax}{1+(a-b)x},\\
\label{A28b}
e^{2\nu}&=&A^{2}D^{2}[1+(a-b)x]^{2n} \nonumber\\ & & \times 
(1+ax)^{2m}\exp\left[-\frac{\alpha\beta x}{2C(a-b)}\right],
\end{eqnarray}
\end{subequations}
where the constants $m$ and $n$ are given by
\begin{eqnarray}
\label{A29}
m &=& \frac{4\alpha b-2k(1+\alpha)}{8b},\nonumber\\
\label{A30}
n &=& \frac{1}{8bC(a-b)^{2}}\nonumber\\ & & \times
\left[a^{2}C(2k(1+\alpha)-4\alpha b) \right.\nonumber\\ & & \left.
-abC(5k(1+\alpha)-2b(1+5\alpha))\right.\nonumber\\ & & \left. 
+b^{2}(3kC(1+\alpha)-2bC(1+3\alpha)+2\beta)\right].\nonumber
\end{eqnarray}
This case corresponds to a charged anisotropic sphere with a linear equation of state. This particular model was first found by Thirukkanesh and Maharaj \cite{hh21}.

If we set $l=0$ in $(\ref{A10b})$ then the term $la^{3}x^{3}$ in the electric field is no longer present. The models metric potentials then become
\begin{subequations}
\label{A31}
\begin{eqnarray}   
\label{A31a}
e^{2\lambda}&=&\frac{1+ax}{1+(a-b)x},\\
\label{A31b}
e^{2\nu}&=&A^{2}D^{2}[1+(a-b)x]^{2n}(1+ax)^{2m}\nonumber\\ & & \times
\exp\left[-\frac{ax[Cs(1+\alpha)+2\beta]}{4C(a-b)}\right],
\end{eqnarray}
\end{subequations}
where the constants $m$ and $n$ are equivalent to 
\begin{eqnarray}
\label{A32}
m &=& \frac{4\alpha b-(1+\alpha)(s+2k)}{8b},\nonumber\\
\label{A33}
n &=& \frac{1}{8bC(a-b)^{2}}\nonumber\\ && \times
\left[a^{2}C((1+\alpha)(s+2k)-4\alpha b) \right.\nonumber\\& &\left.
-abC(5k(1+\alpha)-2b(1+5\alpha))\right.\nonumber\\
& & \left.+b^{2}(3kC(1+\alpha)-2bC(1+3\alpha)+2\beta)\right].\nonumber
\end{eqnarray}
This case describes a model of a charged anisotropic sphere with a barotropic linear equation of state. The charged solution $(\ref{A31})$ was first found by Mafa Takisa and Maharaj \cite{hh22}. Our new class of solutions is therefore a generalisation of previously known models. It arises because of the additional term with the constant $l$ added in the electric field intensity. 

\section{Physical features\label{sec5}}
\begin{figure}[ht]
\centering
\vspace{0.1cm}
\includegraphics[width=8.1cm,height=6.1cm]{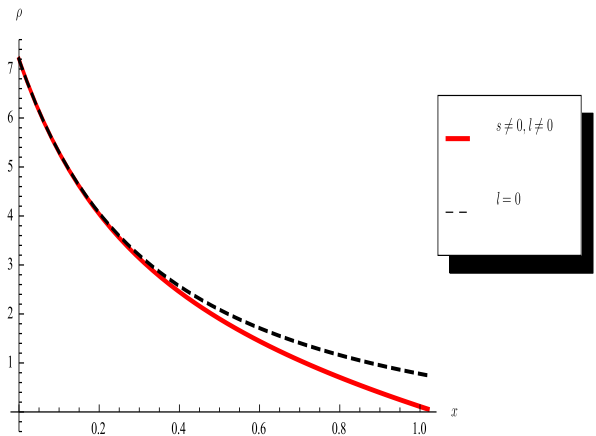}  
\caption{Energy density.}
\label{label:1}
\end{figure}

\begin{figure}[ht]
\centering
\vspace{0.1cm}
 \includegraphics[width=8.1cm,height=6.1cm]{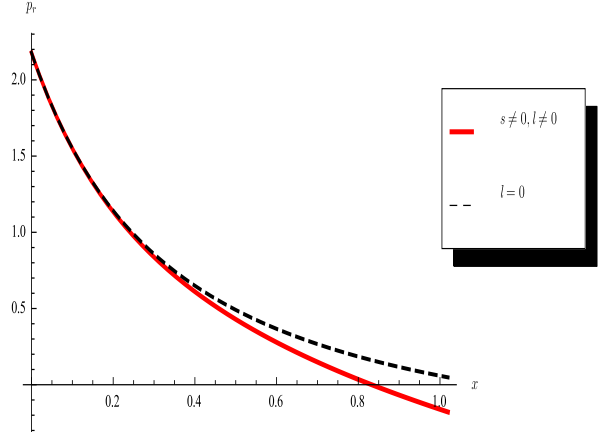}  \caption{Radial pressure.}
 \label{label:2}
\end{figure}

\begin{figure}[ht]
\centering
\vspace{0.1cm}
 \includegraphics[width=8.1cm,height=6.1cm]{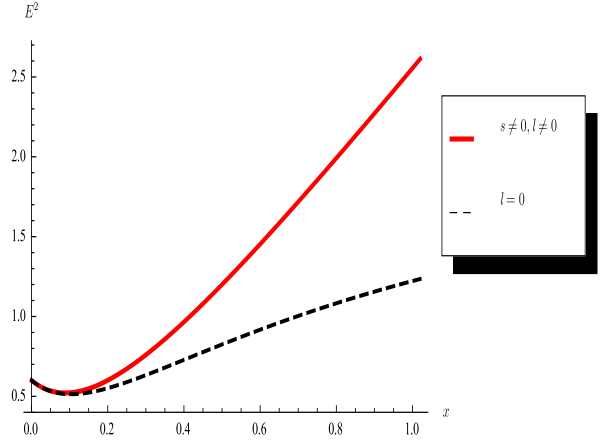}  \caption{Electric field intensity.}
\label{label:3}
\end{figure}

\begin{figure}[ht]
\centering
\vspace{0.1cm}
\includegraphics[width=8.1cm,height=6.1cm]{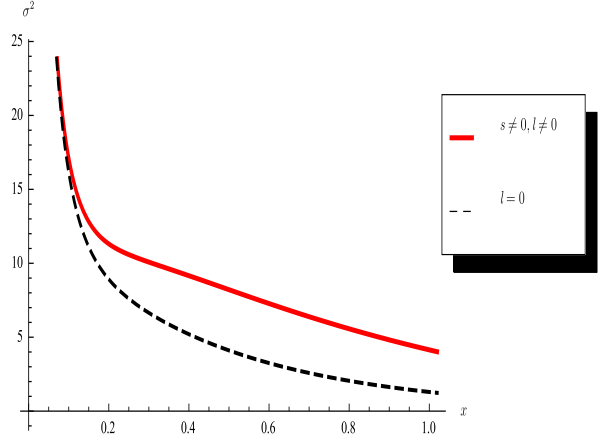}  \caption{Charge density.}
\label{label:4}
\end{figure}

\begin{figure}[ht]
\centering
\vspace{0.1cm}
\includegraphics[width=8.1cm,height=6.1cm]{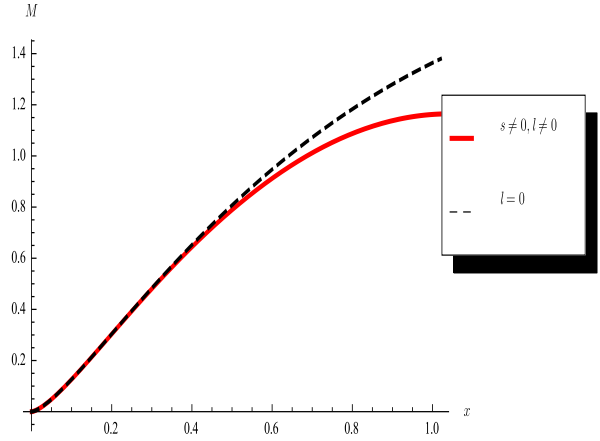}  \caption{Mass function.}
\label{label:5}
\end{figure}

In this work, we have presented a new general model of a relativistic astrophysical star, and integrated a differential equation of first order from the Maxwell-Einstein system of field equations. We can show that this solution is physically reasonable. We utilise the exact solution obtained in Sect.~\ref{g1}. when $a\neq b$ for a graphical analysis. The software package Mathematica (Wolfram \cite{hh25}) was used to generate plots for the matter variables. We made the choices $a=2$, $b=2.5$, $\alpha=0.33$, $\beta=0.198$, $C=1$, $l=1.5$, $s=2.5$, $k=0$ for the various parameters. We have made choices of the parameters similar to that in the analysis of Mafa Takisa and Maharaj \cite{hh22} so that we can be consistent with their analysis. We generated graphical plots for the matter variables, where Figure ~\ref{label:1} shows that the energy density $\rho$ is positive, finite and strictly decreasing. The radial pressure $p_{r}$ in Figure ~\ref{label:2}, expressed in terms of $\rho$ in accordance with the equation of state, follows a similar evolution as the density energy. In Figure ~\ref{label:3} we observe the electric field $E$ initially decreases and then increases after reaching a minimum. The charge density $\sigma$ is a regular and decreasing function in Figure ~\ref{label:4}. In Figure ~\ref{label:5}, the mass function $M$ is continuous, finite and strictly increasing. In the graphs plotted the dashed line corresponds to the case $l=0$, and the solid line corresponds to the case $l\neq 0$, $s\neq 0$. The overall profiles of the matter variables $\rho$, $p_{r}$, $E$, $\sigma$ and $M$ in our investigation are similar to those generated by Mafa Takisa and Maharaj \cite{hh22} when $l=0$. However in the presence of the additional term in the electric field including the parameter $l\neq 0$, we observe that there is some change in the gradients of the respective profiles. The density $\rho$, radial pressure $p_{r}$ and mass $M$, have lower values when $l\neq 0$. The electric field and charge density have higher values when $l\neq 0$. The effect of the parameter $l$ is consequently to enhance and strengthen the effect of the electromagnetic field in the relativistic star.

\newpage
\section{Stellar masses\label{sec6}}
In this section we generate stellar masses which are consistent with the results of Sharma and Maharaj \cite{hh26}, Thirukkanesh and Maharaj \cite{hh21}, and Mafa Takisa and Maharaj \cite{hh22}. Therefore the solutions found in this paper may be applied to realistic stellar astronomical bodies. We use the transformation
\begin{eqnarray}
\label{A34}
\tilde{a}=aR^2 , ~~\tilde{b}=bR^2 ,~~ \tilde{\beta}=\beta R^2 , \nonumber\\ 
~~ \tilde{k}=kR^2 ,~~ \tilde{s}=sR^2 ,~~ \tilde{l}=lR^2.
\end{eqnarray}
This is the same transformation, extended to include the additional parameter $l$, that was applied by Mafa Takisa and Maharaj \cite{hh22}. When $l=0$ in $(\ref{A34})$ we obtain previous results. With the transformation $(\ref{A34})$, and setting $C=1$, we can write the energy density $(\ref{A25c})$ and the mass function $(\ref{A26})$ as
\begin{subequations}
\label{A35}
\begin{eqnarray}
\label{A35a}
\rho&=&\frac{(2\tilde{b}-\tilde{k})(3+\tilde{a}y)-\tilde{s}\tilde{a}^{2}y^{2}-\tilde{l}\tilde{a}^{3}y^{3}}{2R^{2}(1+\tilde{a}y)^{2}} ,\\
\label{A35b}
M&=&\frac{\tilde{l}r(-6\tilde{a}^{3}y^{3}+14\tilde{a}^{2}y^{2}-70\tilde{a}y-105)}{120\tilde{a}(1+\tilde{a}y)}\nonumber\\ & & + \frac{r^{3}(-3\tilde{k}+6\tilde{b}+5\tilde{s})}{12R^{2}(1+\tilde{a}y)}+\frac{\tilde{s}r(15-2\tilde{a}^{2}y^{2})    }{24\tilde{a}(1+\tilde{a}y)}\nonumber\\ & & + \frac{R(7\tilde{l}-5\tilde{s})}{8a^{3/2}}\arctan(\sqrt{\tilde{a}y}),
\end{eqnarray}
\end{subequations}
where $y=\frac{r^{2}}{R^{2}}$.

If we set $\tilde{l}=0$, $\tilde{s}\neq 0$, $\tilde{k}\neq 0$, with $E\neq 0$, then $(\ref{A35})$ yields
\begin{subequations}
\label{A36}
\begin{eqnarray}
\label{A36a}
\rho&=&\frac{(2\tilde{b}-\tilde{k})(3+\tilde{a}y)-\tilde{s}\tilde{a}^{2}y^{2}}{2R^{2}(1+\tilde{a}y)^{2}},\\
\label{A36b}
M&=&\frac{r^{3}(-3\tilde{k}+6\tilde{b}+5\tilde{s})}{12R^{2}(1+\tilde{a}y)}+\frac{\tilde{s}r(15-2\tilde{a}^{2}y^{2})}{24\tilde{a}(1+\tilde{a}y)}\nonumber\\ & & - \frac{5\tilde{s}R}{8a^{3/2}}\arctan(\sqrt{\tilde{a}y}).
\end{eqnarray}
\end{subequations}
These expressions above are related to the solutions in Sect.~\ref{sec4} when $l=0$, found by Mafa Takisa and Maharaj \cite{hh22}. If we set $\tilde{l}=0$, $\tilde{s}=0$, $\tilde{k}\neq 0$ with $E\neq 0$ then we obtain
\begin{subequations}
\label{A37}
\begin{eqnarray}
\label{A37a}
\rho&=&\frac{(2\tilde{b}-\tilde{k})(3+\tilde{a}y)}{2R^{2}(1+\tilde{a}y)^{2}},\\
\label{A37b}
M&=&\frac{r^{3}(2\tilde{b}-\tilde{k})}{4R^{2}(1+\tilde{a}y)},
\end{eqnarray}
\end{subequations}
which are related to the solutions in Sect.~\ref{sec4} when $s=l=0$ found by Thirukkanesh and Maharaj \cite{hh21}. If we set $\tilde{l}=0, \tilde{s}=0, \tilde{k}=0$ with $E=0$ then we obtain
\begin{subequations}
\label{A38}
\begin{eqnarray}
\label{A38a}
\rho&=&\frac{\tilde{b}(3+\tilde{a}y)}{R^{2}(1+\tilde{a}y)^{2}} ,\\
\label{A38b}
M&=&\frac{\tilde{b}r^{3}}{2R^{2}(1+\tilde{a}y)},
\end{eqnarray}
\end{subequations}
which are the solutions attributed to Sharma and Maharaj \cite{hh26}.

We have produced a variety of stellar masses for particular choices of the parameters $\tilde{a}$ and $\tilde{b}$. These are presented in the following tables: Table \ref{table:1} (Stellar masses with $l=0, s=0$), Table \ref{table:2} (Stellar masses with $l=0, s\neq 0$), Table \ref{table:3} (Stellar masses with $l\neq 0, s\neq 0$) and Table \ref{table:4} (Comparative masses). We have set $r=7.07$ km and $R=43.245$ km which are the values used by Dey et al. \cite{hh27, hh28}. For the electric field we have set $\tilde{k}=37.403$,  $\tilde{s}=0.137$ and  $\tilde{l}=0.111$.

For $l=0$, $s=0$ and $k=0$ our model reduces to an uncharged stellar body as that in Sharma and Maharaj \cite{hh26} in terms of the masses generated. When $l=0$, $s=0$ and $k\neq 0$, the masses obtained are similar to that found by Thirukkanesh and Maharaj \cite{hh21} for a charged relativistic star model. Note that for $l=0$, $s\neq 0$ and $k\neq 0$ we have a charged relativistic stellar body, and the stellar masses are consistent with the values generated by Mafa Takisa and Maharaj \cite{hh22}. To distinguish between the various cases in the presence of the electrical field we let $k\neq 0$, $l=0$, $s=0$ in Table \ref{table:1}, $k\neq 0$, $l=0$, $s\neq 0$ in Table \ref{table:2} and $k\neq 0$, $l\neq 0$, $s\neq 0$ in Table \ref{table:3}. We have also included the case for $k=0$, in all tables, so that we can compare with uncharged masses. In Table \ref{table:4} we gather all results and provide a comparative table. It is interesting to observe the effect of the electric field on the masses generated. The introduction of the new parameter $l$ does not appear to appreciably change the mass for the parameter values chosen; a different set of parameters can produce a different profile for the mass. It is important to observe that our results generalise previous treatments and regain the stellar masses of charged and uncharged objects reported in earlier investigations. In particular observe that the parameter values $\tilde{b}=54.34$ and $\tilde{a}=53.340$ correspond to the analysis of Dey  et al. \cite{hh27, hh28} for an equation of state with strange matter. The values obtained are consistent with results of observed masses for the X-ray binary pulsar SAX J1808.4-3658. Consequently our new charged general relativistic solutions of the Einstein-Maxwell system are of astrophysical importance.

\newpage
\begin{table}
\caption{Stellar masses with $l=0, s=0$}
\label{table:1}
\centering 
\begin{tabular}{|c|c|c|c|c|c|c|}
\hline  &  & $M(M_{\bigodot})$ & $M(M_{\bigodot})$ \\
$\tilde{b}$ & $\tilde{a}$ &   $k=s=l=0$ &
$l=0, s=0$\\
& &  $(E=0)$ & $(E\neq 0)$\\\hline
30 ~~~~&    23.681 & 1.175& 0.4426\\
40 ~~~~&    36.346 & 1.298 & 0.6911\\
50 ~~~~&    48.307 & 1.396 & 0.8738\\
54.34  &    53.340 & 1.434 & 0.9399\\
60 ~~~~&    59.788 & 1.478 & 1.0169\\
70~~~~ &    70.920 & 1.547 & 1.1333\\
80 ~~~~&    81.786 & 1.607 & 1.2308\\
90 ~~~~&    92.442 & 1.659 & 1.3141\\
100~~~~ &   102.929& 1.706 & 1.3864\\
\hline
\end{tabular}
\end{table}

\begin{table}
\centering \caption{Stellar masses with $l=0, s\neq 0$}
\label{table:2}
\begin{tabular}{|c|c|c|c|c|c|c|}
\hline  &  & $M(M_{\bigodot})$ & $M(M_{\bigodot})$ \\
$\tilde{b}$ & $\tilde{a}$ &   $k=s=l=0$ &
$l=0, s\neq0$\\
& &  $(E=0)$ & $(E\neq 0)$\\\hline
30 ~~~~&    23.681 & 1.175& 0.4425\\
40 ~~~~&    36.346 & 1.298 & 0.6909 \\
50 ~~~~&    48.307 & 1.396 & 0.8736\\
54.34  &    53.340 & 1.434 & 0.9396\\
60 ~~~~&    59.788 & 1.478 & 1.0165\\
70~~~~ &    70.920 & 1.547 & 1.1329\\
80 ~~~~&    81.786 & 1.607 & 1.2303\\
90 ~~~~&    92.442 & 1.659 & 1.3136\\
100~~~~ &   102.929& 1.706 & 1.3860 \\
\hline
\end{tabular}
\end{table}

\begin{table}
\caption{Stellar masses with $l\neq 0, s\neq 0$}
\label{table:3}
\centering 
\begin{tabular}{|c|c|c|c|c|c|c|}
\hline  &  & $M(M_{\bigodot})$ & $M(M_{\bigodot})$ \\
$\tilde{b}$ & $\tilde{a}$ &   $k=s=l=0$ &
$l\neq 0, s\neq 0$\\
& &  $(E=0)$ & $(E\neq 0)$\\\hline
30 ~~~~&    23.681 & 1.176 & 0.4424\\
40 ~~~~&    36.346 & 1.298 & 0.6907\\
50 ~~~~&    48.307 & 1.396 & 0.8733\\
54.34  &    53.340 & 1.434 & 0.9393\\
60 ~~~~&    59.788 & 1.478 & 1.0162\\
70~~~~ &    70.920 & 1.547 & 1.1324\\
80 ~~~~&    81.786 & 1.607 & 1.2297\\
90 ~~~~&    92.442 & 1.659 & 1.3129\\
100~~~~ &   102.929& 1.706 & 1.3850\\
\hline
\end{tabular}
\end{table}
\begin{table*}[htbp]
\caption{Comparative masses} 
\label{table:4}
\centering
\begin{tabular}{|c|c|c|c|c|c|c|}
\hline &  & $M(M_{\bigodot})$ & $M(M_{\bigodot})$
& $M(M_{\bigodot})$ & $M(M_{\bigodot})$ \\
 $\tilde{b}$ & $\tilde{a}$ &  $k=s=l=0$ &
$l=0, s=0$ & $l=0, s\neq0$ & ($l\neq 0, s\neq0$)\\
& &  $(E=0)$ & $(E\neq 0)$ & $(E\neq 0)$ & $(E\neq 0)$\\\hline
30 ~~~~&    23.681 & 1.175& 0.4426 & 0.4425 & 0.4424\\
40 ~~~~&    36.346 & 1.298 & 0.6911 & 0.6909 & 0.6907\\
50 ~~~~&    48.307 & 1.396 & 0.8738 & 0.8736 & 0.8733\\
54.34  &    53.340 & 1.434 & 0.9399 & 0.9396 & 0.9393\\
60 ~~~~&    59.788 & 1.478 & 1.0169 & 1.0165 & 1.0162\\
70~~~~ &    70.920 & 1.547 & 1.1333 & 1.1329 & 1.1324\\
80 ~~~~&    81.786 & 1.607 & 1.2308 & 1.2303 & 1.2297\\
90 ~~~~&    92.442 & 1.659 & 1.3141 & 1.3136 & 1.3129\\
100~~~~ &   102.929& 1.706 & 1.3864 & 1.3860 & 1.3850\\
\hline
\end{tabular}
\label{tab:LPer}
\end{table*}
\clearpage
\newpage

\section{ Discussion\label{sec7}}
In this paper we presented new exact solutions to the Einstein-Maxwell system of equations with a barotropic equation of state. The form of the quark equation of state chosen between the radial pressure and the energy density is linear. This new class of solutions to the Einstein-Maxwell system of equations obtained in this work is physically reasonable in an astrophysical context and serves to model relativistic stars. We have demonstrated this by graphically analysing the matter variables, and also generating various tables for the mass function for different values of the parameters in the electric field. Our general class of solutions contain the earlier models of Mafa Takisa and Maharaj \cite{hh22} and Thirukkanesh and Maharaj \cite{hh21} in the presence of charge. The models of Dey  et al. \cite{hh27, hh28} and Sharma and Maharaj \cite{hh26} are regained in the appropriate limit for vanishing charge. The stellar masses generated are consistent with earlier investigations and we regain the stellar mass of the object SAX J$1808.4-3658$ as $1.434M_{\bigodot}$. Our new family of solutions solutions is a further indication of the richness of the Einstein-Maxwell system of equations and their link to astrophysical applications; we studied the physical features and plotted the matter and electrical variables. A comparative table of masses for uncharged and charged matter was established which corresponds to observed astronomical objects. The parameter $l$ does not appear to appreciably change the mass for the parameter values chosen but a different set of parameters can give a different profile for the mass.

%
%
%
\section*{ACKNOWLEDGEMENTS}
DKM thanks the National Research Foundation and the University of KwaZulu-Natal for financial support. SDM acknowledges that this work is based upon research supported by the South African Research Chair Initiative of the Department of Science and Technology and the National Research Foundation.
%

\end{document}